\documentclass[aps,prb,reprint,floatfix,superscriptaddress]{revtex4-1}

\usepackage{amsmath,amsfonts} %
\usepackage{wasysym} %
\usepackage[acronym]{glossaries} %
\usepackage{graphicx} %
\usepackage{float}
\usepackage{dcolumn} %
\usepackage{bm} %
\usepackage{ifpdf} %
\usepackage{color} %
\usepackage{nicefrac} %
\usepackage{natbib} %
\usepackage[byname]{smartref} %
\usepackage[breaklinks, %
colorlinks, %
linkcolor=blue, %
citecolor=blue, %
urlcolor=blue]{hyperref} %
\usepackage[table]{xcolor} %
\usepackage{braket} %
\usepackage{mathrsfs}
\usepackage{subfigure}
\usepackage{csquotes} 
\usepackage{comment} 
\usepackage{ulem}
\newcommand\norm[1]{\left\lVert#1\right\rVert}
\newcommand{\eq}[1]{Eq.~(\ref{#1})} %
\begin{document}

\title{Controlling energy conservation in quantum dynamics with independently moving basis functions: Application to Multi-Configuration Ehrenfest}

\author{Mina Asaad}
  \affiliation{Department of Physical and Environmental Sciences,
  University of Toronto Scarborough, Toronto, Ontario, M1C 1A4,
  Canada} %
  \affiliation{Chemical Physics Theory Group, Department of Chemistry,
  University of Toronto, Toronto, Ontario, M5S 3H6, Canada} %
\author{Lo\"ic Joubert-Doriol}
  \affiliation{Univ Gustave Eiffel, Univ Paris Est Creteil, CNRS, UMR 8208, MSME, F-77454 Marne-la-Vall\'ee, France}
\author{Artur F. Izmaylov}
  \affiliation{Department of Physical and Environmental Sciences,
  University of Toronto Scarborough, Toronto, Ontario, M1C 1A4,
  Canada} %
  \affiliation{Chemical Physics Theory Group, Department of Chemistry,
  University of Toronto, Toronto, Ontario, M5S 3H6, Canada} %

\begin{abstract}
Application of the time-dependent variational principle to a linear combination of frozen-width Gaussians describing the nuclear wavefunction provides a formalism where the total energy is conserved. The computational downside of this formalism is that trajectories of individual Gaussians are solutions of a coupled system of differential equations, limiting implementation to serial propagation algorithms. To allow for parallelization and acceleration of the computation, independent trajectories based on simplified equations of motion were suggested. Unfortunately, within practical realizations involving  finite Gaussian bases, this simplification leads to breaking the energy conservation. We offer a solution for this problem by using Lagrange multipliers to ensure the energy and norm conservation regardless of basis function trajectories or basis completeness. We illustrate our approach within the Multi-Configuration Ehrenfest method considering a linear vibronic coupling model.
\end{abstract}

\maketitle

\section{Introduction}

Numerical methods for studying non-adiabatic dynamics require solving the molecular time-dependent Schr\"odinger equation. Grid-based methods such as Multi-Configurational Time-Dependent Hartree (MCTDH), \cite{meyer2009multidimensional, meyer1990multi, beck2000multiconfiguration} excel at the quantum description of dynamics, however system size is limited to ~10 atoms. Limitations arise from exponential scaling of the grid size with respect to the number of degrees of freedom. Recent developments of MCTDH go beyond traditional implementation to a few hundred degrees of freedom, however, generating the nuclear Hamiltonian potential remains the limiting factor.\cite{richings2018mctdh,lasorne2007direct} 
MCTDH requires parametrized potential energy surfaces (PES) or solving the time-independent Schr\"odinger equation for the electronic degrees of freedom at each grid point. Ideally, only points where the nuclear wavepacket travels would be calculated as quantum chemistry calculations are the most computationally intensive step. This is the idea behind on-the-fly dynamics and unfortunately, it is not directly applicable to MCTDH. Instead, new developments in MCTDH fit the PES using kernel ridge regression in an attempt to propagate the wavefunction on-the-fly. \cite{richings2018mctdh}

The work of Heller \cite{heller1981frozen} paved an alternative approach to the grid-based methods using a basis of multidimensional frozen-width Gaussian wave-packets (FW-GWPs).\cite{ben2002ab,curchod2018ab,LoicRevJPCA:2018} The success of FW-GWP methods such as G-MCTDH\cite{ meyer2009multidimensional, burghardt1999approaches, worth2003full}, Ab Initio Multiple Spawning (AIMS),\cite{ben1998nonadiabatic,ben2002ab,curchod2018ab,neville2016beyond,neville2018ultrafast} Multi-Configurational Ehrenfest (MCE),\cite{shalashilin2009quantum, shalashilin2010nonadiabatic,makhov2014ab} and Variational Multi-Configurational Gaussians (vMCG) \cite{worth2004novel, richings2015quantum} is due to the localized shape of Gaussian functions. Thanks to their local character, FW-GWPs decrease the region where PES must be calculated since it can be approximated with a low order polynomial expansion. These polynomial terms are integrated analytically further reducing the computational overhead. Additionally, FW-GWP based methods eliminate the problem of exponential scaling of computational cost with the system size for short-time dynamics. The number of FW-GWP parameters scale linearly with the number of nuclear degrees of freedom. Owing to locality of the nuclear wavefunction in ultra-fast processes, it can be described with only a few FW-GWPs that are parametrized by their centre positions and momenta.

Outstanding concerns faced by most FW-GWP methods are convergence for long-term dynamics and numerical stability. \cite{richings2015quantum, polyak2019direct} For long-term dynamics, FW-GWPs spread out in the Hilbert space, where their locality acts as a double edged sword. An initial dense basis of FW-GWP becomes sparse and inadequate to represent the wavefunction's evolution. Spawning and cloning methods adaptively increase the basis set to address the convergence problem.\cite{ben2002ab,makhov2014ab,IzmaylovPTS:2013,izmaylov2017quantum} However, when FW-GWP basis functions coalesce during propagation, rank deficient matrices are formed. This gives rise to numerical instabilities in the equations of motion and may hinder the propagation of the wavefunction. To correct for numerical instabilities, other methods aim to dynamically remove linear dependencies between FW-GWPs that lead to rank deficiencies. \cite{habershon2012linear}

All approaches based on moving FW-GWPs can be separated in two groups depending on whether the variational time-dependent principle (TDVP) is used for determining FW-GWP evolutions or not. For example, vMCG uses TDVP for 
determining trajectories of FW-GWP, while AIMS and MCE involve classical and Ehrenfest equations of motions, respectively, for the same purpose. There are several computational advantages in removing the TDVP propagation of  FW-GWPs. First, this eliminates matrix inversion instabilities arising in propagation of TDVP's coupled basis trajectories.\cite{richings2015quantum}
Second, independent FW-GWP trajectories can be computed efficiently using parallelization. The disadvantage of independent trajectories is that energy conservation is not guaranteed when the basis set is incomplete. Usually, to observe energy deviations one needs to have FW-GWPs that can transfer population between each other and then evolve following each its own path. Such a scenario frequently can take place in description of electronic population transfer or nuclear wavepacket interference.

Considering closed systems, an energy conserving propagation scheme is necessary to contain wave-packet propagation within energy permitted regions. Furthermore, treating realistic systems with a complete basis set is not computationally feasible. Therefore, it is imperative to consider the case of an incomplete basis when developing propagation schemes. In this paper we develop techniques to maintain energy conservation within independently moving basis functions.  \hyperref[section:Theory]{Section II} illustrates the energy derivative is generally non-zero. We propose the McLachlan Variational Principle \cite{mclachlan1964variational} with a penalty function and Lagrange multipliers to ensure energy and norm conservation. Finally in  \hyperref[section:Numerical_Examples]{Sec. III}, the results between MCE and the proposed variant, Constrained Variational Ehrenfest (CVE) are compared for a two-state, two-dimensional, linear vibronic coupling model.

\section{Theory}
\label{section:Theory}
\subsection{Constrained Variational Ehrenfest}

We start with the following general ansatz for non-adiabatic dynamics
\begin{align}
\Psi(\mathbf{r},\mathbf{R},t) &= \sum_{k} C_k(t) \Phi_k(\mathbf{r},\mathbf{R},t),
\end{align}
where $\Phi_k(\mathbf{r},\mathbf{R},t)$ are time-dependent basis functions of electronic ($\mathbf{r}$) and nuclear ($\mathbf R$) degrees of freedom, and $C_k(t)$ are time-dependent complex coefficients. To find the time evolution of $\Psi$ for an arbitrary $\dot \Phi_k$, we solve the time-dependent Schr\"odinger equation. Varying $\dot C_k$ parameters to minimize the norm of the error,
\begin{align}
\label{eqn:err}
	\ket{{\rm err}} = \ket{\dot \Psi + i\hat H\Psi},
\end{align}
the McLachlan Variational Principle leads to a coupled system of differential equations for $C_k$\cite{worth2004novel, richings2015quantum}
\begin{align}
\label{eqn:old_Coeff}
	\mathbf{\dot C} &= \mathbf{S^{-1}}\left[ -i\mathbf{H} -\bm{\tau}\right]\mathbf{C},
\end{align}
with matrices defined as
\begin{align}
	S_{kl} &= \left<\Phi_k|\Phi_l\right> \\
	H_{kl} &= \left<\Phi_k|\hat H|\Phi_l\right> \\
	\tau_{kl} &= \left<\Phi_k|\dot \Phi_l\right>,
\end{align}
and $\hat H$ is the system Hamiltonian. The problem of using an arbitrary time evolution for the basis $\Phi_l$ with Eq. (\ref{eqn:old_Coeff}) is a non-zero energy derivative
\begin{align}
\label{eqn:Edot}
	\dot E = 2 Re\left\{\sum_l \left< \Psi \left|\hat H(\hat 1 - \hat P)\right| \dot \Phi_l \right>C_l\right\},
\end{align}
where the projector operator is
\begin{align}
\label{eqn:proj}
\hat P = \sum_{kl}\ket{\Phi_k}[\mathbf S^{-1}]_{kl}\bra{\Phi_l}.
\end{align}

There are two key characteristics of the exact solution: conservation of energy and the wavefunction norm.
To constrain the energy variation and to keep the wavefunction norm fixed we use the following Lagrangian 
\begin{align}
\label{eqn:lagrangian}
\mathscr{L} &= \norm{\ket{{\rm err}}}^2+ \beta(\dot E_{\rm CVE})^2 + 2\lambda Re\left<\Psi |\dot \Psi \right>,
\end{align}
where $\beta (\dot E_{\rm CVE})^2$ is a penalty function and $\lambda$ is the Lagrange multiplier for the norm.
One may argue that minizing the Lagrangian with a non-zero penalty function would imply finding a solution corresponding to a non-zero error. However, the error defined in Eq. (\ref{eqn:err}) is an approximation when the basis does not span the entire Hilbert space. Therefore, by adding the penalty function we optimize wavefunction parameters to provide dynamics with the smallest error for a given energy constraint. 

Varying $\dot C_k$ and $\lambda$ parameters to minimize $\mathscr{L}$ we obtain 
\begin{align}
	\label{eqn:Var_Ck}
	0 &= \left< \Phi_k| {\rm err} \right> + 2\beta \dot E_{\rm CVE} \left<\Phi_k |\hat H \Psi \right> + \lambda \left<\Phi_k | \Psi \right> \\
	\label{eqn:Var_norm}
	0 &= 2Re\left<\Psi |\dot \Psi \right>.
\end{align}
The solution of the above system of equations for $\mathbf{\dot C}$ is derived in \hyperref[section:Coeff_EOM]{Appendix A}, and can be written as the following matrix expression
\begin{align}
\label{eqn:new_Coeff}
	\mathbf{\dot C} &= \mathbf{S^{-1}}\left[ -i\mathbf{H} -\bm{\tau} - 2\beta \dot{E}_{\rm CVE}\left( \mathbf{H}-E\mathbf{S}\right)\right]\mathbf{C}.
\end{align}

Using the above energy constrained equation for the coefficients, the corresponding energy derivative labeled $\dot E_{\rm CVE}$ derived in \hyperref[section:Ener_var]{Appendix B} is
\begin{align}
\label{eqn:Edot_CVE}
	\dot E_{\rm CVE} =\frac{2 Re\left\{\sum_{l} \left< \Psi \left|\hat H(\hat 1 - \hat P)\right| \dot \Phi_l \right>C_l\right\} }{1 + 4\beta \left(\mathbf{C^\dagger H S^{-1}H C -} E^2 \right)}.
\end{align}
In the above equation, $\beta$ is a free parameter arbitrarily assigned a value to limit energy variation. In the limit where $\beta \rightarrow \infty$, $\dot E_{\rm CVE} \rightarrow 0$ and is later referred to as CVE. However, $\beta$ may also be defined as a time dependent parameter such that the energy derivative does not exceed a set threshold value ($\dot E_{\rm thr}$). By doing so, the user has direct control over the allowed energy fluctuation. This is done by re-arranging Eq. (\ref{eqn:Edot_CVE}) to solve for an adaptive $\beta$ at each time step before solving  Eq. (\ref{eqn:new_Coeff}),
\begin{align}
\label{eqn:Adapt_Beta}
	\beta(t) &=\frac{2 Re\left\{\sum_l\left< \Psi \left|\hat H(\hat 1 - \hat P)\right| \dot \Phi_l \right>C_l\right\} - \dot E_{\rm thr} }{ 4\dot E_{\rm thr} \left(\mathbf{C^\dagger H S^{-1}H C - }E^2 \right)}.
\end{align}
It is shown in \hyperref[section:Norm_cons]{Appendix C} that the wavefunction norm is conserved if coefficients are evolved according to \eq{eqn:new_Coeff}.

Yet, besides the energy and norm conservation, there is another property of the exact propagation that can be violated by approximate schemes, it is unitarity. The exact propagator is unitary, and thus it maintains a constant overlap between two distinct states described by $\mathbf{C}$ and $\mathbf{C'}$. To check whether the new method is unitary, the time derivative of the overlap between two states is calculated,
\begin{align}
\label{eqn:Unitary}
	\frac{\partial}{\partial t} \left( \mathbf{C^\dagger S C'} \right) &= \mathbf{C^\dagger} \left[- 2\beta \dot{E}_{\rm CVE}\left( \mathbf{H}-E\mathbf{S}\right)\right]\mathbf{C'} \notag \\
	&+ \mathbf{C^\dagger} \left[- 2\beta \dot{E}_{\rm CVE}'\left( \mathbf{H}-E'\mathbf{S}\right)\right]\mathbf{C'},
\end{align}
where $ \dot{E}_{\rm CVE}$ and  $\dot{E}_{\rm CVE}'$ calculated with $\mathbf{C}$ and $\mathbf{C'}$, respectively. 
Equation  (\ref{eqn:Unitary}) demonstrates that our new constraint on energy and norm does not preserve the inner product unless $\beta\dot{E}_{CVE}$ and $\beta\dot{E}_{CVE}'$ vanish exactly. According to \eq{eqn:Edot_CVE}, $\dot {E}_{CVE} \rightarrow 0$ can be achieved in the $\beta \rightarrow \infty$ limit, however, product $\beta\dot{E}_{CVE}$ goes to non-zero constant in that limit unless the energy is conserved as in Eq. (\ref{eqn:Edot}). Therefore, CVE is unitary only when the energy constraint is not required to ensure the energy conservation. Comparing Eqs. (\ref{eqn:old_Coeff}) and (\ref{eqn:new_Coeff}) highlights the addition of the $\beta$-term associated with the penalty function as the culprit of non-unitarity problem. 
The $\beta$-term is hermitian while the remaining terms are anti-hermitian. Anti-hermitian matrices lead to a unitary evolution due to cancellations when summed with its hermitian conjugate in Eq. (\ref{eqn:Unitary})

\subsection{Application to Multi-Configurational Ehrenfest}

To illustrate our method for modelling quantum dynamics with independently moving basis functions, it is applied to MCE. The MCE wavefunction is expanded over a basis consisting of products of electronic states, expanded over diabatic states $\psi^{(s)}(\mathbf{r})$, and FW-GWPs ($g_k$)~\cite{shalashilin2009quantum}
\begin{align}
\Phi_k(\mathbf{r},\mathbf{R},t) = g_k(\mathbf{R};\mathbf{z_k}) \left( \sum_s a_k^{(s)}(t) \psi^{(s)}(\mathbf{r}) \right),
\end{align}
where parameters $\mathbf{z_k}=\sqrt{\frac{\omega}{2}}\mathbf{x}_{\mathbf{k}} + \frac{i}{\sqrt{2\omega}}\mathbf{p_k}$ encode average positions $\mathbf{x}_k$ and momenta $\mathbf{p}_k$ of the FW-GWPs.
Gaussian amplitudes $a_k^{(s)}$ determine weighted average to obtain an Ehrenfest potential surface on which the FW-GWPs evolve classically. The MCE equations of motion are \cite{shalashilin2009quantum, shalashilin2010nonadiabatic}
\begin{align}
\label{eqn:adot}
	\mathbf{\dot a_k} &= i\left[i\frac{\mathbf{z_k^\dagger\dot z_k} - \mathbf{\dot z_k^\dagger z_k}}{2}\ -\mathbf{H_k}\right] \mathbf{a_k}\\
\label{eqn:zdot}
	i \dot z_{k_\kappa} &= \frac{\partial H_k^{Ehr}}{\partial z_{k_\kappa}^*},
\end{align}
where $\kappa$ refers to FW-GWP degrees of freedom,
the matrix $\mathbf{H_k}$ in Eq. (\ref{eqn:adot}) is defined as 
\begin{align}
{H_k^{ss'}} &= \braket{g_k \psi^{(s)}|\hat H |g_k \psi^{(s')}},
\end{align}
and the Ehrenfest Hamiltonian is 
\begin{align}
	 H_k^{Ehr} = \braket{\Phi_k | \hat H | \Phi_k}.
\end{align} 

Finally, the equations of motion for the basis coefficients ($C_k$) are obtained as in Eq. (\ref{eqn:new_Coeff}). Energy is exactly conserved when $\beta \rightarrow \infty$. In the limit $\beta \rightarrow 0$, Eq. (\ref{eqn:new_Coeff}) coincides with the original MCE Eq. (\ref{eqn:old_Coeff}). Thus, $\beta$ is a continuous parameter connecting CVE and MCE.

\section{Numerical Examples}
\label{section:Numerical_Examples}
To assess the performance of CVE we consider a two-dimensional, two-state, linear vibronic coupling (2D-LVC) model. This model is chosen for its simplicity and presence of non-adiabatic effects. The system Hamiltonian in the diabatic representation with mass-weighted coordinates is
\begin{align}
\hat H_{\rm LVC} &= \mathbf{1} \otimes \hat T_N + 
\begin{bmatrix}
\frac{\omega^2}{2}\left[(\hat x-a)^2 + \hat y^2 \right] & c \hat y \\
 c \hat y & \frac{\omega^2}{2}\left[(\hat x+a)^2 + \hat y^2 \right]
\end{bmatrix},
\end{align}
where
\begin{align}
\hat T_N =-\frac{1}{2} \left(\frac{\partial^2}{\partial x^2} +  \frac{\partial^2}{\partial y^2}\right),
\end{align}
and
\begin{align}
\omega=2\text{, } \hspace{5mm} a=1.5 \text{, } \hspace{5mm}c=6.
\end{align}
The initial wavefunction consists of two FW-GWPs centred at the local minima $(-1.5,0)$ with momenta $(0.1,\pm 0.5)$ opposite in the $y$-direction
\begin{align}
	\Psi(\mathbf{r},\mathbf{R},0) \propto \sum_{\pm}g\left(\mathbf{R};[-1.5,0] + \frac{i}{2}[0.1,\pm 0.5]\right) \psi^{(1)}(\mathbf{r}),
\end{align}
it splits as FW-GWPs propagate around both sides of a conical intersection (located at the centre of the coordinate system in the adiabatic representation).

\begin{figure}[H]
    \centering
    \includegraphics[trim={1cm 0 1cm 0},width=\linewidth]{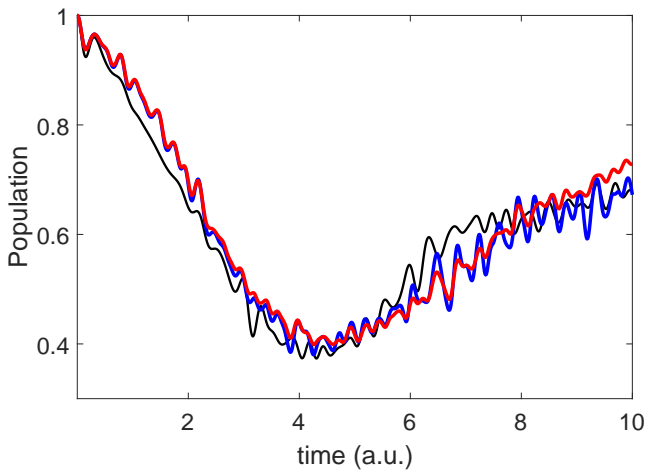}
    \caption{Ground state population for MCE (blue) and CVE (red) with 128 FW-GWP basis, shown in comparison with SOA (black).}
    \label{GS_pop_Conv}
\end{figure}

To demonstrate the negligible effect of $\beta$ in Eq. \ref{eqn:new_Coeff} on population dynamics for a complete basis, and later consider the effects for an incomplete basis, we first need to determine what constitutes a \enquote{complete} basis. This is done by comparing MCE and CVE against split-operator algorithm (SOA) \cite{tannor2007introduction} which is taken to be the exact solution. An indicator for SOA accuracy is energy fluctuation which we define as the ratio of energy change over system energy. The maximum energy variation of SOA is $\sim 10^{-4}$, where 200 points are taken over the interval $(-17,17)$ for each degree of freedom and a time increment of 0.0005 a.u. Our \enquote{complete} basis for MCE and CVE is obtained with 128 FW-GWP basis functions and ground state population is shown in comparison with the exact calculation in Fig. \ref{GS_pop_Conv}. We note the minimal population variation of MCE from CVE as they both converge to the exact results, despite a maximum energy fluctuation for MCE double the system energy.

\begin{figure}[H]
    \centering
    \includegraphics[trim={1cm 0 1cm 0},width=\linewidth]{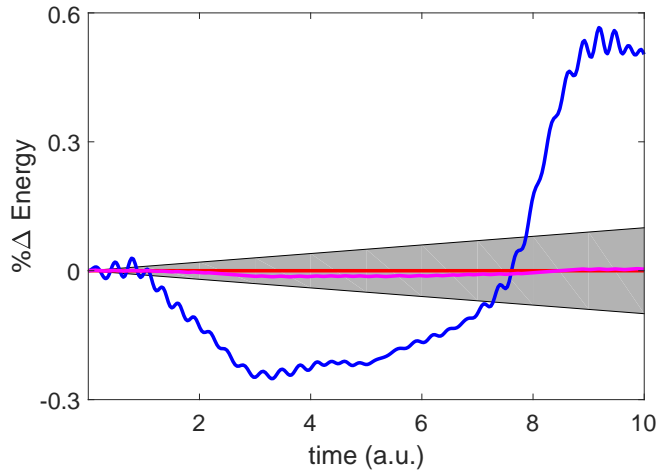}
    \caption{Variation of energy for MCE (blue) shown in comparison to the CVE (red).  Depicted in magenta is the case with adaptive parameter $\beta$, where the shaded region depicts upper and lower bounds for permitted energy variation.}
    \label{Energy}
\end{figure}

However, our primary interest is to study the circumstances where MCE and CVE diverge as a consequence of an incomplete basis. Therefore it is imperative to compare MCE and CVE against vMCG, where the same basis functions are used. Variational Multiconfiguration Gaussian conserves both energy, norm and unitarity and differs from MCE only in how the basis functions move. To simulate an incomplete basis, 6 FW-GWP functions are used instead of 128 previously used to approximate convergence. Figure \ref{Energy} shows CVE maintains energy conservation in the limit where $\dot E_{\rm thr} = 0$ (equivalent to $\beta \rightarrow \infty$), while MCE does not. Furthermore, we see the implementation of Eq. (\ref{eqn:Adapt_Beta}), by setting $\dot E_{\rm thr} = 0.02$ and note the upper bound is linear in time. The results shown in magenta for $\beta (t)$ in the same figure, are well within this boundary. Having succeeded in amending MCE to conserve energy, we consider the significance of energy conservation in ground state population dynamics.

\begin{figure}[H]
    \centering
    \includegraphics[trim={1cm 0 1cm 0},width=\linewidth]{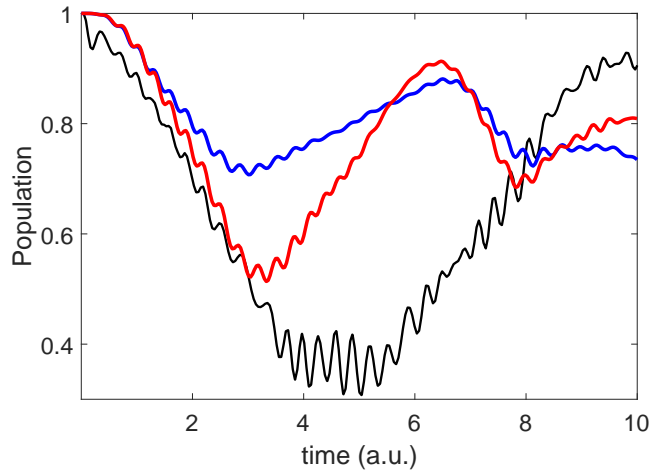}
    \caption{Ground state population for MCE (blue) and CVE (red) with 6 FW-GWP basis, shown in comparison with vMCG results (black).}
    \label{GS_pop}
\end{figure}

The results in Fig.  \ref{GS_pop} present an example where CVE provides a better description for population dynamics in comparison to MCE over short timescales. Naturally, this desired characteristic would be attributed to energy conservation in CVE as system energy affects the spread of the wavefunction. However, since the energy conservation is only a necessary but not sufficient condition for the exact dynamics, CVE does not always outperform MCE in agreement with vMCG population dynamics.

\section{Conclusions}
In this work we derived a method for constraining energy to an arbitrary threshold for independently moving basis functions. 
The method was applied to MCE dynamics for the 2D-LVC model and was referred to as CVE. The CVE method has a scalable parameter that is responsible for the degree of energy conservation ($\beta$). For the model system, tuning the parameter allowed us to limit the relative energy variation to 10\% and by taking the $\beta \rightarrow \infty$ limit to achieve the full energy conservation. Considering differences in electronic population dynamics between the vMCG (full quantum with energy conservation), MCE, and CVE methods, it was found that initial CVE dynamics agrees with that of vMCG for longer than the one produced by MCE. Yet, the full conservation of energy in the $\beta \rightarrow \infty$ limit breaks down 
another property of the exact dynamics, unitarity. Unitarity is conserved in the CVE scheme only for $\beta=0$. Thus, the $\beta$-parameter can be used in CVE to tune the energy-conservation/unitarity trade-off as desired.Our approach is completely general, and this balance between energy conservation and unitarity is in fact more widely applicable to other type of conserved quantities (e.g. momentum or angular momentum conservation for spectrocopic calculation).

Furthermore, the method developed in this paper allows for parallelization of the basis movement without sacrificing energy conservation and can be applied to AIMS and other schemes making use of independently moving basis functions. In addition to its moving basis flexibility, the method allows one to freely choose a convenient electronic basis.

\section*{Appendix A: CVE Coefficient Equation of Motion}
\label{section:Coeff_EOM}
Minimizing the Lagrangian from Eq. (\ref{eqn:lagrangian}) with respect to $C_k$'s, Eq. (\ref{eqn:Var_Ck}) 
can be expressed in the matrix notation as
\begin{align}
\label{eqn:Var_Ck_sim}
	0 &= \mathbf{S}\mathbf{\dot C} + \bm{\tau}\mathbf{C} + i\mathbf{H} \mathbf{C} + 2\beta \dot E_{\rm CVE} \mathbf{H} \mathbf{C} + \lambda \mathbf{S} \mathbf{C}.
\end{align}
Rewriting Eq. (\ref{eqn:Var_Ck_sim}) for $\mathbf{\dot C}$,
\begin{align}
\label{eqn:coeff_step1}
	\mathbf{\dot C} &= \mathbf{S^{-1}}\left[- i\mathbf{H} -\bm{\tau} - 2\beta \dot E_{\rm CVE} \mathbf{H} - \lambda \mathbf{S}\right] \mathbf{C}.
\end{align}
To solve for $\lambda$, we consider $ [ \mathbf{C^\dagger}$ Eq.(\ref{eqn:Var_Ck_sim}) $+$  
Eq. (\ref{eqn:Var_Ck_sim})$^\dagger \mathbf{C}]$, 
\begin{align}
\label{eqn:coeff_step2}
	0 &= \mathbf{C^\dagger}\mathbf{S}\mathbf{\dot C} + \mathbf{C^\dagger}\bm{\tau}\mathbf{C} + i\mathbf{C^\dagger}\mathbf{H} \mathbf{C} + 2\beta \dot E_{\rm CVE} \mathbf{C^\dagger}\mathbf{H} \mathbf{C} + \lambda \mathbf{C^\dagger}\mathbf{S} \mathbf{C} \notag \\
 &+\mathbf{\dot C^\dagger}\mathbf{S}\mathbf{C} + \mathbf{C^\dagger}\bm{\tau^\dagger}\mathbf{C} - i\mathbf{C^\dagger}\mathbf{H} \mathbf{C} + 2\beta \dot E_{\rm CVE} \mathbf{C^\dagger}\mathbf{H} \mathbf{C} + \lambda \mathbf{C^\dagger}\mathbf{S} \mathbf{C}.
\end{align}
Recalling norm conservation from Eq. (\ref{eqn:Var_norm})
\begin{align}
\tag{\ref{eqn:Var_norm} revisited}
	0 &= 2Re\left<\Psi |\dot \Psi \right> \\
	&= \mathbf{C^\dagger}\mathbf{S}\mathbf{\dot C} + \mathbf{C^\dagger}\bm{\tau}\mathbf{C} + \mathbf{\dot C^\dagger}\mathbf{S}\mathbf{C} + \mathbf{C^\dagger}\bm{\tau^\dagger}\mathbf{C},
\end{align}
we solve for $\lambda$ in Eq. (\ref{eqn:coeff_step2})
\begin{align}
\lambda=-2\beta \dot E_{\rm CVE} E.
\end{align}
Substituting $\lambda$ into Eq. (\ref{eqn:coeff_step1}), we arrive at Eq. (\ref{eqn:new_Coeff}),
\begin{align}
\tag{\ref{eqn:new_Coeff} revisited}
	\mathbf{\dot C} &= \mathbf{S^{-1}}\left[ -i\mathbf{H} -\bm{\tau} - 2\beta \dot{E}_{\rm CVE}\left( \mathbf{H}-E\mathbf{S}\right)\right]\mathbf{C}.
\end{align}

\section*{Appendix B: CVE Energy variation}
\label{section:Ener_var}
Time derivative of the average energy for a time-independent Hamiltonian is 
\begin{align}
	\dot E_{\rm CVE} &= \frac{\partial \braket{\Psi|\hat H|\Psi}}{\partial t} \\
	&= 2 Re\left\{\braket{\Psi|\hat H|\dot \Psi}\right\}\\
	&= 2 Re\left\{\sum_l \braket{\Psi|\hat H|\dot \Phi_l}C_l + \braket{\Psi|\hat H|\Phi_l}\dot C_l\right\}.
\end{align}
Substituting $\dot C_l$ from Eq. (\ref{eqn:new_Coeff}),
\begin{align}
	\dot E_{\rm CVE} = 2 Re&\left\{\sum_l \left< \Psi \left|\hat H(\hat 1 - \hat P)\right| \dot \Phi_l \right>C_l \right.\\
	&\left.- 2\beta \dot E_{\rm CVE} \mathbf{C^\dagger H S^{-1}}\left(\mathbf{H}-E\mathbf{S}\right)\mathbf{C}\right\}. \notag
\end{align}
Noting the first term is $\dot E$ from Eq. (\ref{eqn:Edot}) and the second term is real,
\begin{align}
	\dot E_{\rm CVE} = \dot E- 4\beta \dot E_{\rm CVE} \mathbf{C^\dagger H S^{-1}}\left(\mathbf{H}-E\mathbf{S}\right)\mathbf{C}.
\end{align}
Finally, re-arranging for $\dot E_{\rm CVE}$ yields Eq. (\ref{eqn:Edot_CVE}).

\section*{Appendix C: CVE Norm Conservation}
\label{section:Norm_cons}
We demonstrate the wavefunction norm conservation by showing that the derivative of the norm is zero 
at all times for the normalized wavefunction evolving according to Eq. (\ref{eqn:new_Coeff}),

\begin{align}
	\frac{\partial}{\partial t} \braket{\Psi|\Psi} &= \frac{\partial}{\partial t} \mathbf{C^\dagger SC}\\
	&= 2 Re\left\{\mathbf{C^\dagger} \bm{\tau} \mathbf{C} + \mathbf{C^\dagger S \dot C}\right\}\\
	&= 2 Re\left\{\mathbf{C^\dagger} \bm{\tau} \mathbf{C}  \right. \\
	&\hspace{7mm} \left.+ \mathbf{C^\dagger}\left[ -i\mathbf{H} -\bm{\tau} - 2\beta \dot{E}_{\rm CVE}\left( \mathbf{H}-E\mathbf{S}\right)\right]\mathbf{C}\right\} \notag \\
	&= 2 Re\left\{- 2\beta \dot{E}_{\rm CVE}\mathbf{C^\dagger}\left( \mathbf{H}-E\mathbf{S}\right)\mathbf{C}\right\} \\
	&= Re\left\{\mathbf{C^\dagger}\mathbf{S}\mathbf{C} - 1\right\} 4\beta E\dot{E}_{\rm CVE} = 0.
\end{align}

\normalem
\bibliography{CVEArticle}
\end{document}